\input epsf

\documentstyle[draft]{aa}
\def\ltsim{\mathrel{\hbox{\rlap{\hbox{\lower3pt\hbox{$\sim$}}}\raise2pt\hbox{$<$}}}}
\begin{document}
\newcommand{\be}{Be\ }
\newcommand{\etal}{{et al.}\ }
\renewcommand{\dbltopfraction}{.5}
\renewcommand{\topfraction}{.5}
\renewcommand{\dblfloatpagefraction}{0.50}
\renewcommand{\floatpagefraction}{0.55}

\thesaurus{08.03.4; 08.05.2; 08.18.1; 03.20.4; 03.13.2} 
 
\title{A Representative Sample of \be Stars IV:\\
Infrared Photometry and the Continuum Excess}

\author{Lee Howells, I.A.~Steele, John M.~Porter
\and J.~Etherton}                   
                                                            
\institute{Astrophysics Research Institute, Liverpool John Moores University, 
Egerton Wharf, Birkenhead, CH41 1LD, United Kingdom}

\offprints{L.~Howells}
\mail{lxh@astro.livjm.ac.uk}

\date{Received  / Accepted }

\titlerunning{A Representative Sample of \be Stars IV}
\authorrunning{L.~Howells \textit{et al}}
\maketitle 

\begin{abstract} 
We  present infra-red ($JHK$) photometry  of 52 isolated \be stars of spectral types O9--B9 and luminosity classes III--V.  We describe a new method of reduction, enabling separation of interstellar reddening and circumstellar excess.  Using this technique we find that the disc emission makes a maximum contribution to the optical $(B-V)$ colour of a few tenths of a magnitude. We find strong correlations between a range of emission lines (H$\alpha$, Br$\gamma$, Br11, and Br18)  from the Be stars' discs, and the circumstellar continuum excesses. We also find that stellar rotation and disc excess are correlated.

\end{abstract}
\section{Introduction}
Classical \be stars are observed to be rotating close to their break up
speeds (Slettebak \cite{s82}) with a modal value of $\sim0.7v_{crit}$ (Porter \cite{p96}), where  $v_{crit}$ is the stars' critical break-up velocities.
The distribution of circumstellar material giving rise to the emission
features in their spectra (leading to their designation ``e'') is
concentrated in the equatorial plane (\cite{1992Natur.359..808D},
\cite{1994A&A...283L..13Q}).
Emission line profiles, \textit{e.g.} see Dachs \etal (\cite{d86}), Hummel \& Vrancken (\cite{hr2000}) are most easily explained if it is assumed that there are Keplerian motions within a disc. Observations of \be star discs have shown that  some can  vanish and
re-appear in a apparently random fashion with time-scales of
100s of days (\textit{e.g.} \cite{1995PASJ...47..195H}, \cite{1993A&A...274..356H}). 
The major problem in \be star research is to identify the physical
processes which produce and disperse the discs.

Theoretical models which have been presented include wind bi-stability
(\cite{1991A&A...244L...5L}), wind compressed discs (Bjorkman \& Cassinelli \cite{bc93}), and viscous outflowing discs (Lee \etal\cite{1991MNRAS.250..432L}, also see Porter \cite{1999A&A...348..512P}). Each of these models has
problems: the wind models (bi-stability and wind compression) cannot reproduce the
Keplerian rotation in the disc as the wind conserves angular momentum
leading to a rotation law similar to $v_{\phi}\propto 1/r$ (see \cite {1994ApJ...424..887O}).
Viscous outflowing discs successfully reproduce most of the 
attributes of the observed \be star discs (\textit{e.g.} continuum excess Porter \cite{1999A&A...348..512P}, $V/R$ variations in the emission lines \cite{1997A&A...318..548O}), although the
major problem is that the required source of angular momentum to sustain
the disc remains obscure (a tentative suggestion has been made that the
angular momentum source relates to non-radial pulsations, \cite{1986PASP...98...30O}).

Consequently our understanding of \be stars is still incomplete.
Theoretical models can only be constrained, or indeed ruled out, if their
predictions can be confronted with observations of well understood \be star
samples.
Hitherto, a homogeneous data set involving extensive wavelength coverage
across all \be spectral types has been lacking.

In a series of papers we have been addressing that demand by defining
and observing in a homogeneous fashion a representative sample of \be
stars.  The sample contains objects from O9 to B8.5 and of luminosity
classes III (giants) to V (dwarfs), as well as three shell stars.  It
was selected in an attempt to contain several objects that were
typical of each spectral and luminosity class in the above range; it
therefore does {\em not} reflect the spectral and luminosity class
space distribution of \be stars, but only the average properties of
each subclass in temperature and luminosity. A spectral type and
measure of $v \sin (i)$ were derived for each object in the sample and
were presented in Steele \etal(1999, hereafter
\cite{p1}).  In Clark \& Steele (2000, hereafter \cite{p2}) we
presented $K$ band spectroscopy of the sample, and in Steele \& Clark
(2000, hereafter~\cite{p3}) $H$ band spectroscopy. In a forthcoming
paper, Steele \& Negueruela (2001, hereafter \cite{p5}) will
present spectra in the regions of the H$\alpha$ and Paschen series
lines.  

This paper presents infrared $JHK$ photometry of the sample
and a new technique for separating out the interstellar reddening and
circumstellar excess.   We then go on to correlate the derived circumstellar
excess of the objects in the sample their emission line strengths, spectral types and 
rotational velocities.  Throughout this paper we shall be  
using non-parametric correlation analysis (using the Spearman rank correlation coefficient; 
\textit{e.g.} Press \etal \cite{spearman}) to study these relationships. In this 
way we impose no assumptions about the form of the dependence between parameters.

In section \ref{sec:obs} we describe our
observations and how we carried out the preliminary data
reduction. Section \ref{sec:disc} describes the process we have used
to separate the circumstellar excess and interstellar reddening
associated with observations of \be stars.  In section \ref{sec:plot}
we present the correlations we have found between the separated
excesses, reddenings and other observed quantities.  
Finally Section~\ref{sec:conc} presents our conclusions.

\section{The Observations}
\label{sec:obs}
Infrared photometry of the sample was obtained on the nights of
1999 September 28--30 the using 155cm Carlos Sanchez Telescope (TCS) 
equipped with the $CVF$ photometer.  The filters used were
the $JHK$ filters of the TCS system (see Alonso \etal \cite{a98}).  Standard stars
taken from the UKIRT bright standards list (the UKIRT system 
is that defined by the UKIRT standards list, observed by the UKT9 photometer) were observed at 
high and low air masses roughly once per 90 minutes.
The standard methods of photometric data reduction were applied, with
airmass corrections derived on a per filter, per night basis.  Colour
corrections from Alonso \etal (\cite{a98}) and Leggett (\cite{leg92}) 
were used to convert the UKIRT standard magnitudes to the TCS system
for reduction.  After reduction in the TCS system, the final magnitudes
were then converted to the standard CIT system (Elias \etal \cite{e82}), and it is these
magnitudes that we present in Table~\ref{tab:mean}.  For those stars 
observed more than once we give a mean value, with the number
of observations identified in brackets after each entry.  
The original sample contains 58 \be stars from O9$\rightarrow$B8.5, however because of observational 
constraints the sample has been reduced, in this paper, to 52 objects.

\begin{table*}
\begin{tabular}{llcccccccccc}
\multicolumn{1}{c}{Star} &
\multicolumn{1}{c}{Spec. type}&
\multicolumn{1}{c}{EW Na(\AA)}&
\multicolumn{1}{c}{J}&
\multicolumn{1}{c}{$\pm$Jerr}&
\multicolumn{1}{c}{H}&
\multicolumn{1}{c}{$\pm$Herr}&
\multicolumn{1}{c}{K}&
\multicolumn{1}{c}{$\pm$Kerr}&
\multicolumn{1}{c}{(10)}&
\multicolumn{1}{c}{(11)}&
\multicolumn{1}{c}{(12)}\\ \hline 
\\
\object{BD$+28\ 03598$} (2) & O9II       & $-1.18$ &  7.31 &0.01&  7.09 &0.01&  6.99 &0.01& 0.23 & -0.08 &  1.18\\
\object{BD$+27\ 00797$} (3) & B0.5V      &   -     &  9.74 &0.03&  9.44 &0.01&  9.38 &0.01& 0.31 & -0.02 &  1.59\\ 
\object{BD$+57\ 00681$}     & B0.5V      & $-0.00$ &  7.63 &0.01&  7.43 &0.01&  7.39 &0.01& 0.23 & -0.15 &  1.17\\ 
\object{BD$+37\ 03856$}     & B0.5V      & $-0.48$ &  9.47 &0.02&  9.34 &0.01&  9.39 &0.01& 0.21 & -0.23 &  1.11\\ 
\object{BD$+56\ 00484$} (2) & B1V        & $-1.15$ &  8.55 &0.02&  8.34 &0.01&  8.11 &0.01& 0.12 &  0.14 &  0.64\\ 
\object{BD$+55\ 00605$} (3) & B1V        & $-1.16$ &  9.37 &0.05&  9.34 &0.01&  9.37 &0.05& 0.15 & -0.09 &  0.78\\ 
\object{BD$+36\ 03946$}     & B1V        & $-0.71$ &  7.28 &0.01&  7.06 &0.01&  6.81 &0.01& 0.11 &  0.18 &  0.58\\ 
\object{BD$+56\ 00478$} (2) & B1.5V      & $-1.15$ &  7.86 &0.01&  7.66 &0.01&  7.46 &0.01& 0.12 &  0.13 &  0.60\\ 
\object{BD$+56\ 00573$} (2) & B1.5V      & $-0.95$ &  8.46 &0.01&  8.02 &0.01&  7.50 &0.01& 0.17 &  0.39 &  0.89\\ 
\object{BD$+29\ 04453$}     & B1.5V      & $-0.40$ &  7.79 &0.01&  7.68 &0.01&  7.42 &0.01& 0.02 &  0.27 &  0.11\\ 
\object{BD$+45\ 00933$}     & B1.5V      & $-0.93$ &  7.47 &0.01&  7.40 &0.01&  7.37 &0.01& 0.10 & -0.04 &  0.51\\ 
\object{BD$+58\ 02320$}     & B2V        & $-0.01$ &  8.79 &0.01&  8.63 &0.01&  8.49 &0.02& 0.12 &  0.06 &  0.62\\ 
\object{BD$+42\ 01376$} (2) & B2V        & $-0.39$ &  6.99 &0.01&  6.88 &0.01&  6.74 &0.01& 0.06 &  0.12 &  0.31\\ 
\object{BD$+05\ 03704$}     & B2.5V      & $-0.38$ &  6.13 &0.01&  6.21 &0.02&  6.22 &0.02&-0.03 &  0.05 & -0.13\\ 
\object{BD$+47\ 00183$}     & B2.5V      & $-0.19$ &  4.43 &0.03&  4.37 &0.02&  4.22 &0.02& 0.04 &  0.14 &  0.19\\ 
\object{BD$+47\ 00939$}     & B2.5V      & $-0.13$ &  3.93 &0.03&  3.82 &0.02&  3.72 &0.02& 0.08 &  0.07 &  0.41\\ 
\object{BD$+42\ 04538$} (2) & B2.5V      & $-0.89$ &  7.78 &0.02&  7.67 &0.01&  7.49 &0.01& 0.05 &  0.16 &  0.27\\ 
\object{BD$+55\ 00552$} (2) & B4V        & $-0.51$ &  8.31 &0.02&  8.30 &0.02&  8.24 &0.01& 0.01 &  0.07 &  0.07\\ 
\object{BD$+30\ 03227$} (3) & B4V        & $-0.36$ &  6.82 &0.03&  6.88 &0.02&  6.94 &0.02& 0.02 & -0.06 &  0.12\\ 
\object{BD$+50\ 00825$} (2) & B7V        & $-0.23$ &  6.02 &0.01&  5.92 &0.01&  5.91 &0.01& 0.09 & -0.06 &  0.46\\ 
\object{BD$-02\ 05328$}     & B7V        & $-0.14$ &  6.32 &0.03&  6.34 &0.02&  6.33 &0.02&-0.01 &  0.04 & -0.06\\ 
\object{BD$+37\ 00675$}     & B7V        & $-0.32$ &  6.21 &0.03&  6.18 &0.02&  6.15 &0.02& 0.03 &  0.02 &  0.13\\ 
\object{BD$+58\ 00554$} (2) & B7V        & $-0.74$ &  8.53 &0.01&  8.43 &0.01&  8.33 &0.01& 0.05 &  0.06 &  0.27\\ 
\object{CD$-22\ 13183$}     & B7V        & $-0.48$ &  7.23 &0.01&  7.13 &0.01&  7.04 &0.01& 0.05 &  0.05 &  0.27\\ 
\object{BD$-00\ 03543$} (2) & B7V        & $-0.42$ &  6.80 &0.01&  6.79 &0.01&  6.77 &0.01& 0.01 &  0.02 &  0.06\\ 
\object{BD$+27\ 03411$}     & B8V        & $-0.09$ &  5.25 &0.03&  5.33 &0.02&  5.35 &0.02&-0.04 &  0.02 & -0.18\\ 
\object{BD$+50\ 03430$}     & B8V        & $-0.21$ &  7.09 &0.01&  7.08 &0.01&  7.06 &0.01& 0.01 &  0.17 &  0.07\\ 
\object{BD$+55\ 02411$} (3) & B8.5V      & $-0.07$ &  7.39 &0.01&  7.28 &0.01&  7.21 &0.01& 0.05 &  0.00 &  0.27\\ 
\object{BD$+20\ 04449$} (2) & B0III      & $-0.48$ &  8.52 &0.01&  8.61 &0.02&  8.69 &0.01& 0.03 & -0.06 &  0.16\\ 
\object{BD$+56\ 00511$} (2) & B1III      & $-1.29$ &  8.31 &0.01&  8.17 &0.01&  8.09 &0.01& 0.14 & -0.02 &  0.72\\ 
\object{BD$+46\ 00275$}     & B5III      & $-0.89$ &  4.29 &0.03&  4.29 &0.02&  4.29 &0.02& 0.04 & -0.02 &  0.20\\ 
\object{BD$+49\ 00614$} (2) & B5III      & $-0.56$ &  7.68 &0.01&  7.67 &0.01&  7.67 &0.01& 0.04 & -0.02 &  0.23\\ 
\object{BD$-19\ 05036$} (2) & B4III      & $-0.58$ &  7.10 &0.06&  6.94 &0.01&  6.90 &0.01& 0.14 & -0.07 &  0.51\\ 
\object{BD$+51\ 03091$}     & B7III      &     -   &  6.06 &0.03&  6.05 &0.02&  6.05 &0.02& 0.03 & -0.01 &  0.13\\ 
\object{BD$+23\ 01148$} (2) & B2III      & $-0.67$ &  6.97 &0.01&  6.78 &0.01&  6.68 &0.01& 0.17 & -0.03 &  0.89\\ 
\object{CD$-28\ 14778$}     & B2III      & $-0.22$ &  8.51 &0.01&  8.33 &0.01&  8.12 &0.01& 0.10 &  0.15 &  0.54\\ 
\object{BD$-12\ 05132$} (2) & BN0.2III   & $-1.07$ &  7.91 &0.01&  7.69 &0.01&  7.48 &0.02& 0.15 &  0.11 &  0.78\\ 
\object{BD$+47\ 03985$}     & B1-2shell  & $-0.56$ &  5.30 &0.01&  5.21 &0.01&  5.07 &0.01& 0.06 &  0.12 &  0.32\\ 
\object{BD$+43\ 01048$} (3) & B6IIIshell & $-1.29$ &  8.90 &0.02&  8.80 &0.01&  8.71 &0.02& 0.08 &  0.03 &  0.39\\ 
\object{BD$+02\ 03815$}     & B7-8shell  & $-0.72$ &  6.58 &0.03&  6.57 &0.02&  6.53 &0.02&-0.01 &  0.07 & -0.03\\ 
\object{CD$-27\ 11872$}     & B0.5V-III  &  -      &  6.89 &0.01&  6.59 &0.01&  6.28 &0.01& 0.17 &  0.19 &  0.86\\ 
\object{BD$+29\ 03842$} (3) & B1II       & $-1.40$ &  9.10 &0.01&  8.98 &0.02&  8.92 &0.01& 0.14 & -0.07 &  0.73\\ 
\object{BD$+56\ 00493$} (3) & B1V-IV     & $-0.74$ &  9.35 &0.02&  9.27 &0.02&  9.24 &0.02& 0.11 & -0.04 &  0.57\\ 
\object{BD$+27\ 00850$}     & B1.5IV     & $-0.76$ &  8.98 &0.02&  8.91 &0.01&  8.92 &0.02& 0.13 & -0.10 &  0.65\\ 
\object{BD$+56\ 00473$} (2) & B1V-III    & $-0.87$ &  8.04 &0.01&  7.78 &0.01&  7.44 &0.01& 0.11 &  0.13 &  0.57\\ 
\object{BD$-01\ 03834$}     & B2IV       & $-0.39$ &  7.67 &0.03&  7.59 &0.02&  7.40 &0.02& 0.01 &  0.22 &  0.08\\ 
\object{BD$+40\ 01213$} (2) & B2.5IV     & $-0.42$ &  7.50 &0.01&  7.50 &0.01&  7.50 &0.01& 0.03 &  0.01 &  0.17\\ 
\object{BD$+47\ 00857$}     & B4V-IV     &   -     &  4.17 &0.03&  4.04 &0.02&  3.88 &0.02& 0.07 &  0.12 &  0.37\\ 
\object{BD$+21\ 04695$}     & B6III-V    & $-0.32$ &  5.91 &0.03&  5.95 &0.02&  5.97 &0.02& 0.01 & -0.01 &  0.03\\ 
\object{BD$+17\ 04087$}     & B6III-V    & $-0.69$ &  9.75 &0.03&  9.72 &0.02&  9.66 &0.02& 0.02 &  0.07 &  0.08\\ 
\object{CD$-27\ 16010$}     & B8IV       & -       &  4.24 &0.03&  4.25 &0.02&  4.24 &0.02& 0.03 & -0.04 &  0.16\\ 
\object{BD$+31\ 04018$}     & BB1.5V     & $-0.69$ &  6.64 &0.03&  6.51 &0.02&  6.33 &0.02&-0.07 &  0.39 & -0.36\\ 
\hline	      
\end{tabular} 
\caption{\label{tab:mean}Our full data sample showing respectively the object, with multiple observations in brackets and spectral type. The Na $D_2$ $5890$\AA\ line EW, $JHK$ photometry post reduction with their associated errors, the results of our separation procedure as $E(H-K)_{is}$ (column 10) \& $E(H-K)_{cs}$  (column 11) and the IR colour $(H-K)$ converted to an equivalent optical $(B-V)$ colour as $E(B-V)_{{(H-K)}_{is}}$  (column 12)}
\end{table*}

\section{Methodology - Separating the Interstellar Reddening and Circumstellar Excess}
\label{sec:disc}

\begin{figure}
\vspace*{7cm}
\includegraphics{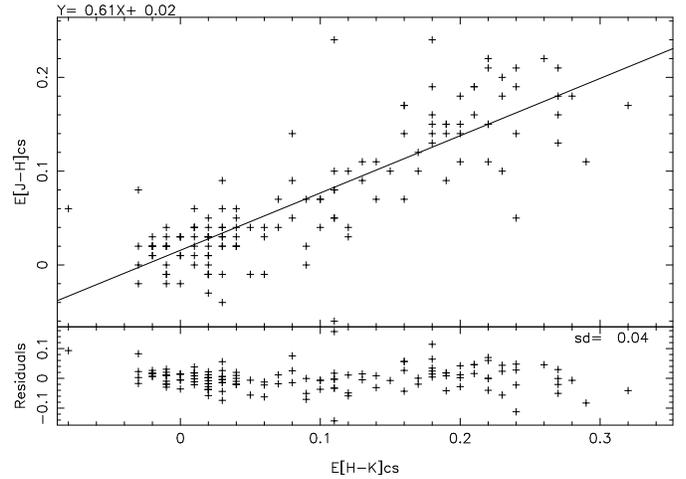}
\caption{Data extracted from Dougherty \etal (\cite{d94}).  The ratio $\beta$ was calculated by means of a least squares fit, shown above as a solid line. Note that ``sd'' is the r.m.s. deviation of the residuals about the fitted line in the ordinate axis.} 
\label{fig:dough}
\end{figure}

The flux from a star is reduced by interstellar 
extinction by a factor ($\exp[-\tau_{ext}(\lambda)]$),
where $\tau_{ext}(\lambda)$ is the extinction optical 
depth.  In general, if we have knowledge of the spectral type of the
object (and hence its intrinsic colour) and the observed colours,
then we can remove extinction effects from data using an interstellar extinction law (\textit{e.g.} Rieke \& Lebofsky \cite{rb85}).  
However \be stars are well known to exhibit an infrared continuum excess,
caused by free-free and free-bound emission within the disc, 
as well as the usual interstellar reddening (\textit{e.g.} Gehrz \etal \cite{g74}).  
At first sight it appears not to be possible to separate the 
interstellar and circumstellar components 
using infrared photometry alone.   However by using the 
fact that the spectral indices of the two components are different,
we find that a deconvolution is possible as follows:

The observed colour, $(M_{\lambda_1}-M_{\lambda_2})_{obs}$, 
of a \be star consists of three components - the star's intrinsic colour, $(M_{\lambda_1}-M_{\lambda_2})_{0}$, the excess due to circumstellar material,
$E(M_{\lambda_1}-M{\lambda_2})_{cs}$ and the interstellar reddening,
$E(M_{\lambda_1}-M_{\lambda_2})_{is}$.  Using our $JHK$ filters we
can construct two observed colours:

\begin{eqnarray}
(J-H)_{obs}=(J-H)_{0}+E(J-H)_{cs}+E(J-H)_{is}
\label{eq:unsol1}
\end{eqnarray} 
\begin{eqnarray}
(H-K)_{obs}=(H-K)_{0}+E(H-K)_{cs}+E(H-K)_{is}.
\label{eq:unsol2}
\end{eqnarray} 

A unique solution for these equations is possible by assuming a 
universal interstellar reddening law of the form:

\begin{eqnarray}
E(J-H)_{is}&=\alpha E(H-K)_{is},
\label{eq:assump1}
\end{eqnarray} 
and also by assuming that the colours of the disc around the \be star 
can be related in a similar fashion (see, for example Figure 5 from Dougherty \etal \cite{d94}, re-plotted, with a least squares fit applied  as our Figure~\ref{fig:dough}):
\begin{eqnarray}
E(J-H)_{cs}&=\beta E(H-K)_{cs}.
\label{eq:assump2}
\end{eqnarray} 

The value of $\alpha$ may be simply derived from the interstellar extinction law 
of Rieke \& Lebofsky (\cite{rb85}), 
giving $\alpha=1.7\pm0.1$.  To derive $\beta$, we use the
circumstellar excesses of Be stars measured by Dougherty \etal (\cite{d94})
who de-redden their photometry based on a combination of the reddening
free Geneva system parameters X and Y, and the strength of the interstellar
2200 {\AA} feature in IUE spectra.  
A least squares fit to the data, giving a $\chi_{reduced}^2=1.02$,
presented in their Figure 5 gives  $\beta=0.61\pm0.02$ (see Figure \ref{fig:dough}).   We note that the r.m.s. deviation in the ordinate direction of the graph, $E[J-H]_{cs}$ vs $E[H-K]_{cs}$ is $sd=0.04$, while that for the graph $E[H-K]_{cs}$ vs $E[J-H]_{cs}$ is $sd=0.05$. we are therefore confident that a 1D minimisation is sufficient.   We note also that a Spearman rank correlation test gives a Spearman rank coefficient, $r=0.8$, when applied to this data, implying a high correlation between the circumstellar excess colours.

By combining equations 1 to 4 we are able to analytically solve to 
separate the interstellar and circumstellar 
components.  We find:

\begin{eqnarray}
\label{eq:sol1}
E(\!H\!\!-\!\!K\!)_{is}=\frac{(\!H\!\!-\!\!K\!)_{obs}\!\!-\!\!\frac{1}{\beta}(\!J\!\!-\!\!H\!)_{obs}\!\!-\!\!(\!H\!\!-\!\!K\!)_{0}\!\!+\!\!\frac{1}{\beta}(\!J\!\!-\!\!H\!)_{0}}{(1\!\!-\!\!\frac{\alpha}{\beta})}\\
\label{eq:sol2}
E(\!H\!\!-\!\!K\!)_{cs}=\frac{(\!H\!\!-\!\!K\!)_{obs}\!\!-\!\!\frac{1}{\alpha}(\!J\!\!-\!\!H\!)_{obs}\!\!-\!\!(\!H\!\!-\!\!K\!)_{0}\!\!+\!\!\frac{1}{\alpha}(\!J\!\!-\!\!H\!)_{0}}{(\!1\!\!-\!\!\frac{\beta}{\alpha})}\\
\label{eq:sol3}
E(\!J\!\!-\!\!H\!)_{is}=\frac{(\!J\!\!-\!\!H\!)_{obs}\!\!-\!\!\frac{1}{\beta}(\!H\!\!-\!\!K\!)_{obs}\!\!-\!\!(\!J\!\!-\!\!H\!)_{0}\!\!+\!\!\frac{1}{\beta}(\!H\!\!-\!\!K\!)_{0}}{(\!1\!\!-\!\!\frac{\beta}{\alpha}\!)}\\
\label{eq:sol4}
E(\!J\!\!-\!\!H\!)_{cs}=\frac{(\!J\!\!-\!\!H\!)_{obs}\!\!-\!\!\frac{1}{\alpha}(\!H\!\!-\!\!K\!)_{obs}\!\!-\!\!(\!J\!\!-\!\!H\!)_{0}\!\!+\!\!\frac{1}{\alpha}(\!H\!\!-\!\!K\!)_{0}}{(\!1\!\!-\!\!\frac{\alpha}{\beta}\!)}
\end{eqnarray} 

The intrinsic colours $(H-K)_0$ and $(J-H)_0$ come from Koornneef (\cite{k83}).
 The solutions of equations~\ref{eq:sol1}--\ref{eq:sol2}  are tabulated in Table~\ref{tab:mean}.  We present only the $(H-K)$ solutions because the $(J-H)$ results are not independent, the ratios $\alpha$ and $\beta$ relating the two. Colour excesses for $(J-H)$ can be simply calculated using  equations~\ref{eq:sol3}--\ref{eq:sol4}.

\begin{figure*}
\vspace*{6.5cm}
\includegraphics{h2419f2a.ps}
\includegraphics{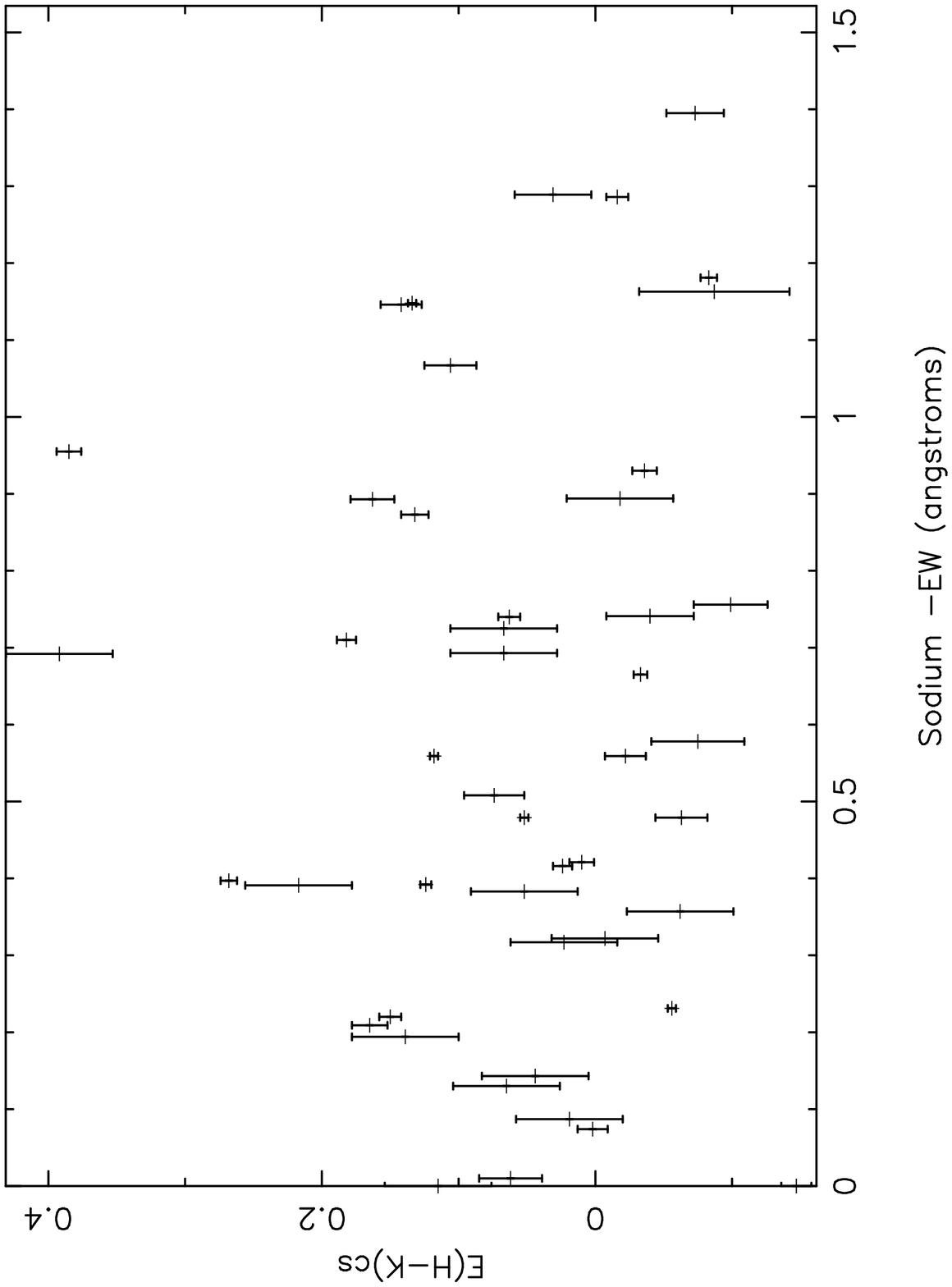}
\caption{Interstellar extinction versus Na $D_2$ $5890$\AA\ line EW (left panel), where the fitted line is a least squares fit weighted to the ordinate axis errors, and versus circumstellar excess (right panel). Note that as expected no correlation  between the \textit{interstellar Na EW} and the \textit{circumstellar} excess is present.}
\label{fig:sodium}
\end{figure*}

The errors generated from our calculations are twofold, (i) random errors from
our observational data and the intrinsic colours, which enables us to
construct errors and quantify the scatter and (ii) systematic errors from 
the ratios $\alpha$ and $\beta$, which shift the calculated best fit lines to their upper and lower extremities.  We calculate a systematic error of $\sim5\%$ in $E(H-K)_{cs}$ and $\sim4\%$ in $E(H-K)_{is}$.

In order to test our de-reddening procedure, we compare the
measured interstellar reddening to an independent measure
of the same quantity.  For this we use 
equivalent width\footnote{ Note that in this paper we will employ the convention that \textit{positive} equivalent widths indicate \textit{emission} features.} (EW) of the interstellar sodium $D_2$ $5890$\AA\ line, listed in column 3 of Table~\ref{tab:mean}. We note that there is an error of 10\% on the Na EW.
This was measured from the red optical spectra of the sample
(see \cite{p5}) using the {\sc figaro}
routine {\sc abline}. In Figure~\ref{fig:sodium} 
we plot the EW of this line against 
our derived interstellar reddening and circumstellar excess.
As expected there appears to be a correlation with 
$E(H-K)_{is}$ although not 
with $E(H-K)_{cs}$. To quantify 
this we performed non-parametric correlation 
tests (Spearman rank).  The results for all such tests carried out in this
paper are presented in Table \ref{tab:spear}.
 We note here that Spearman rank correlation confidences are normally compared with a critical correlation coefficient, $r_{s}$, which imply a significance level for the correlation. we list this significance level for each test in Table \ref{tab:spear}.  However we have also
chosen to express our results as a standard deviation ($\sigma$) measure (confidence level) to allow easy comparison with 
parametric tests.  Implicit in this is the assumption that repeated tests of similar samples would find
a normal distribution of the derived correlation coefficients. To derive this confidence level we used the one-tailed $r_{s}$ lookup tables of \cite{1996QJRAS..37..519W} to find the significance level and then the one-tailed normal distribution lookup tables of \cite{1979QJRAS..20..138W} to find the confidence levels. Therefore we also list in Table \ref{tab:spear} the confidence level of each test.
The positive correlation between sodium EW  and the interstellar 
extinction is confirmed at a $>4.5\sigma$ confidence level while
any correlation between sodium and $E(H-K)_{cs}$
is at a confidence level of less than $1\sigma$.   This result gives us 
confidence that
our method does indeed separate the interstellar and
circumstellar components of the infrared excess.

\begin{figure*}
\vspace*{6.5cm}
\includegraphics{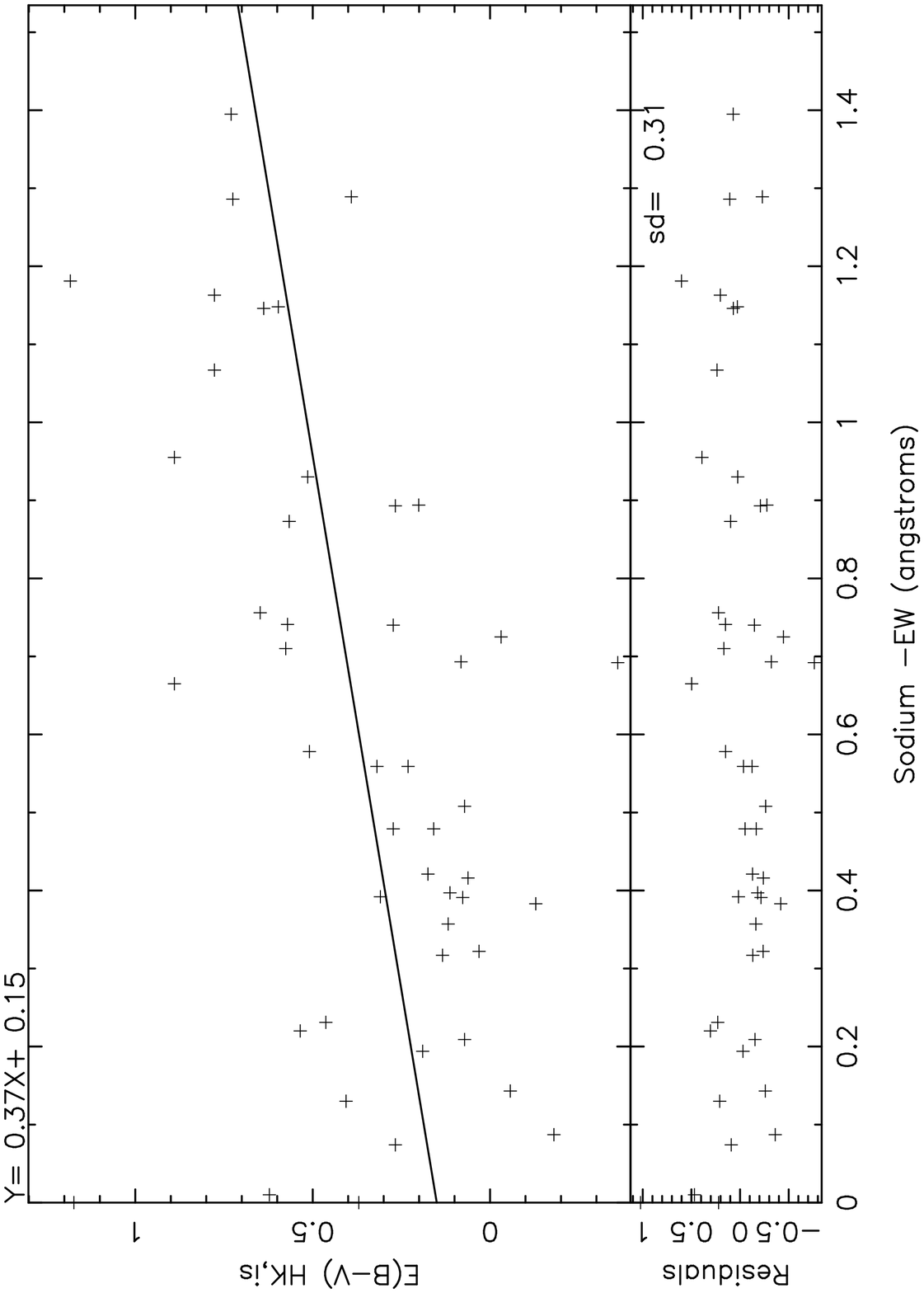}
\includegraphics{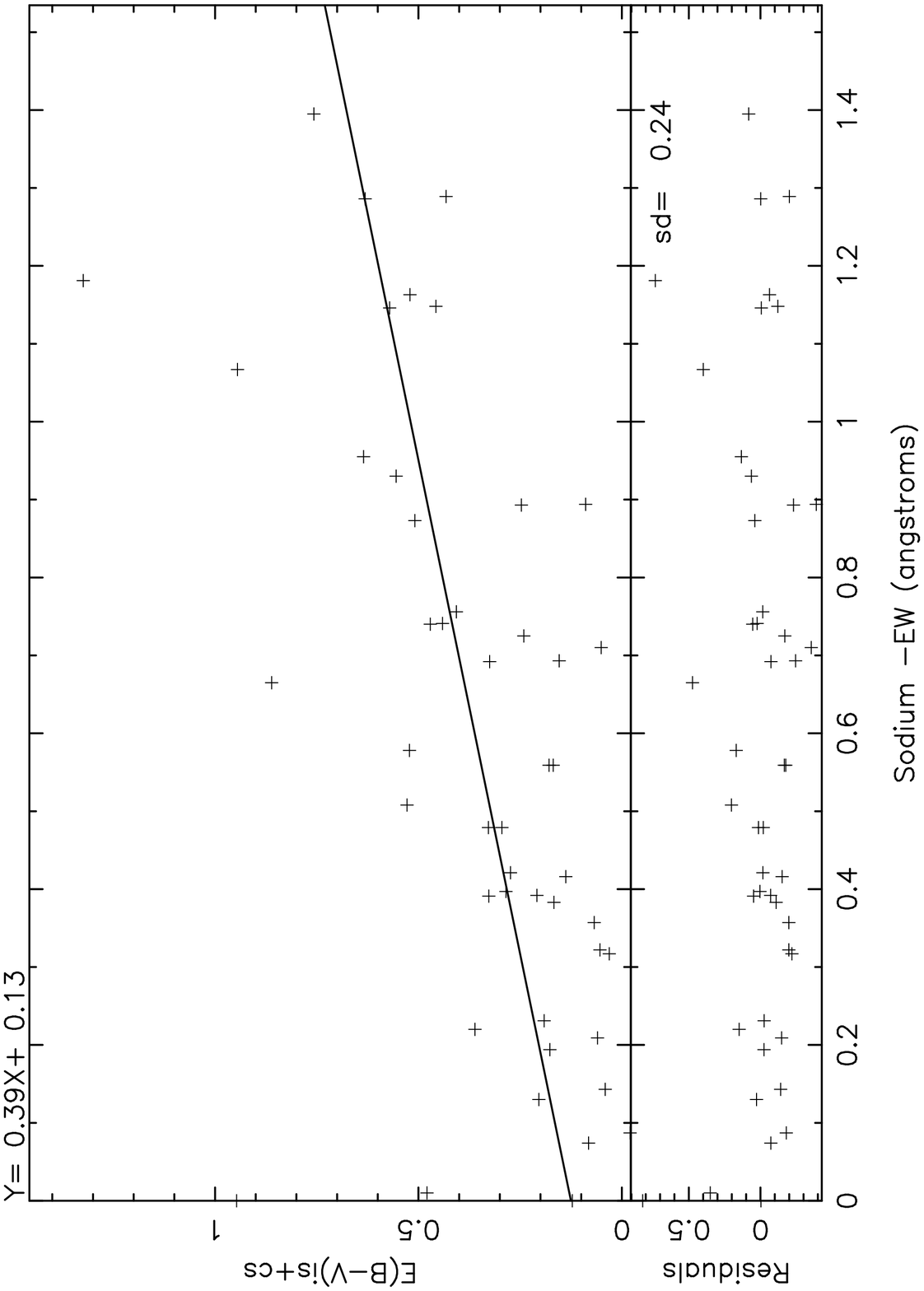}
\caption{(Left panel) a plot of $E(B-V)_{{(H-K)}_{is}}$, the IR $(H-K)$ interstellar reddening, converted to an optical $(B-V)$ colour versus sodium EW(\AA). (Right Panel) a plot of $(B-V)_{is+cs}$ versus  the Na $D_2$ $5890$\AA\ line EW. The fitted lines  are least squares fits minimised in the ordinate axis.}
\label{fig:sods}
\end{figure*}


\begin{figure}[p]
\vspace*{6.5cm}
\includegraphics{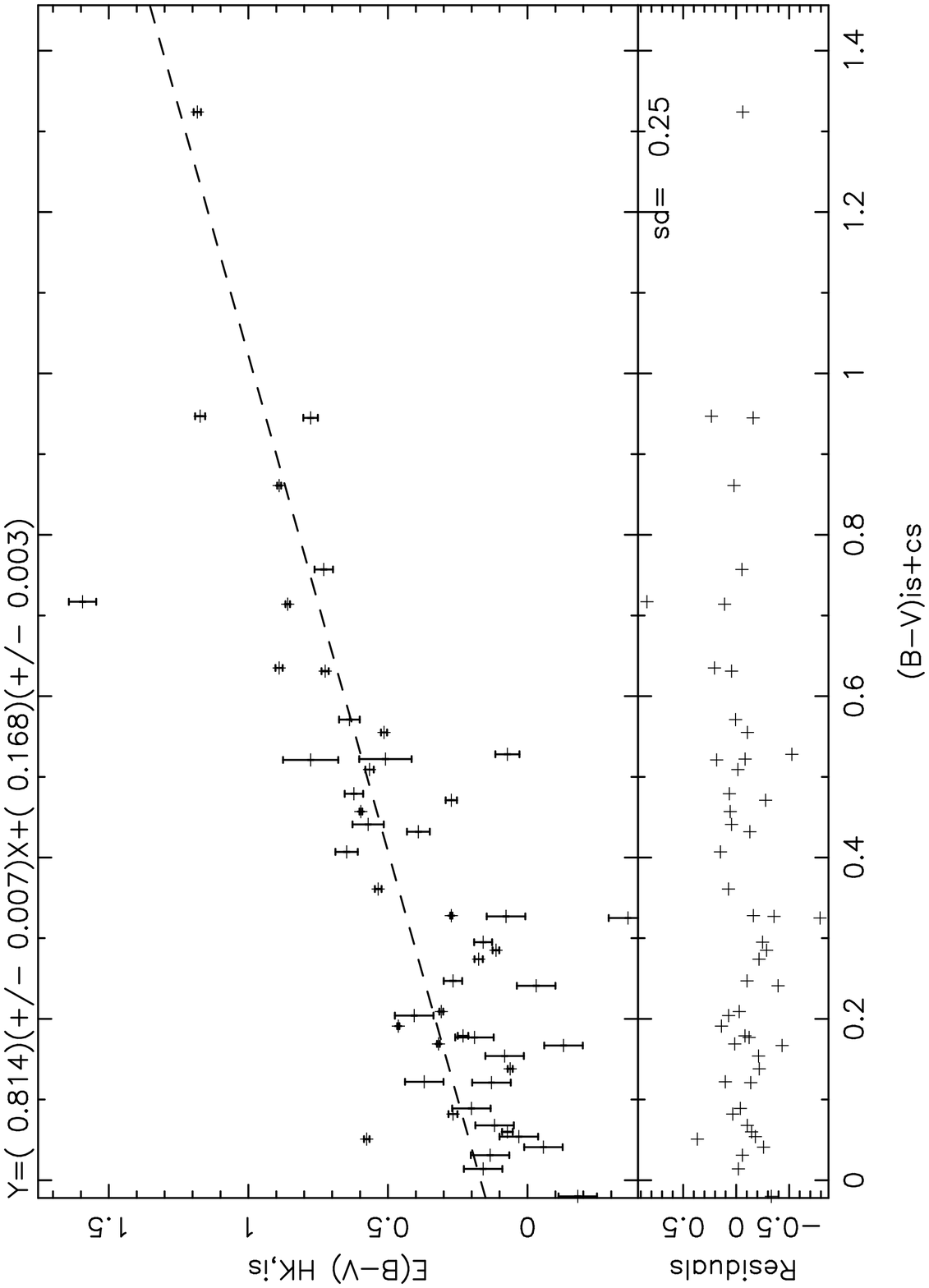}
\caption{A plot of $E(B-V)_{{(H-K)}_{is}}$ (the IR $(H-K)$ interstellar reddening converted to an optical $(B-V)$  colour) versus the optical $(B-V)_{is+cs}$ colour ,  which incorporates both interstellar reddening and circumstellar excess.}
\label{fig:bvhkis}
\end{figure}

To quantify  the strength of any optical  circumstellar excess in our sample we convert our 
IR interstellar excesses to equivalent optical data using our 
adopted interstellar extinction law of Rieke \& Lebosky (\cite{rb85}). The interstellar excess converted 
from an $(H-K)$ colour to a $(B-V)$ equivalent colour is denoted by $E(B-V)_{{(H-K)}_{is}}$. This is plotted 
against $E(B-V)_{cs+is}$, \textit{i.e.} incorporating both interstellar reddening and circumstellar excess 
(see Figure \ref{fig:bvhkis}),  where $\Delta E(B-V)_{{(H-K)}_{is}}\gg\Delta E(B-V)_{is+cs}$ and 
so it is $E(B-V)_{{(H-K)}_{is}}$ that has been minimised. $E(B-V)_{cs+is}$ is derived from historical 
observational data (see \cite{p1}) and the intrinsic $(B-V)$ colours of B stars (Cramer \cite{c84}).  
An independent test of our de-reddening procedure may now be carried out if we assume a negligible circumstellar 
excess for the optical $(B-V)$ colour:  The colour-colour plot should produce a one-to-one correlation if the 
assumption of zero optical excess is true. A correlation is again obvious ($r=0.74$), and we  note that no 
significant offset between the two measures of reddening is apparent. 

This implies that the assumption of negligible optical circumstellar 
excess appears to be reasonable at the level of $<$0.17 magnitudes, 
(the intercept of Figure \ref{fig:bvhkis}).  There is also a systematic error (as described above) of 0.2mags 
associated with the plot in the ordinate direction. This implies boundary
 conditions of $-0.03<E(B-V)_{cs}<0.37$ magnitudes.    A similar result was 
found by Dachs, Kiehling \& Engels (\cite{d88}) who find that the maximum contribution of 
circumstellar envelopes to observed $(B-V)$ colours in \be stars amounts 
to $E(B-V)_{cs}\sim0.1$ magnitudes.  

 In the light of this result (negligible optical circumstellar excess) it would be interesting
to determine which method (optical colours, infrared colours or sodium equivalent width) gives a better estimate of
the interstellar reddening to Be stars.    
The Spearman rank correlation coefficient of $E(B-V)_{is+cs}$ versus the sodium EW (see Figure~\ref{fig:sods}, right panel) is $r=0.56$.
For $E(B-V)_{(H-K)_{is}}$ versus sodium EW (Figure~\ref{fig:sods}, left panel) the Spearman rank correlation coefficient is $r=0.45$.
However the greatest
correlation is between $E(B-V)_{is+cs}$ and $E(B-V)_{{(H-K)}_{is}}$ (see Figure~\ref{fig:bvhkis})  with $r=0.74$.  In other words it appears that
both the traditional optical and our new infrared method are more reliable than the sodium equivalent width
for determining the interstellar reddening to Be stars.  In the sections that follow we prefer to use
our new method, as it is based on data taken closer in time (within a few years) 
to the spectroscopic data than the optical data (over 30 years in many cases).


\section{Results} 
\label{sec:plot}
In this section we plot the separated $E(H-K)_{cs}$ against the equivalent widths of various emission features from the spectra presented in Papers II, III and IV and test to see if the quantities are related by using a Spearman rank test which is non-parametric. Where we believe a linear correlation exists we calculate best fit lines using a least squares fit weighted to errors in the ordinate axis.  For each of these \textit{linear} cases we display the calculated equations above each plot and the standard deviation, in the ordinate direction, about the fitted line in the residuals section of the plot. These values are also recorded in Table~\ref{tab:spear}.

We removed the underlying photospheric absorption by combining our spectra with values of corresponding B star absorption lines tabulated in Hanson \etal (\cite{h96}) and Hanson \etal (\cite{h98}). 
 
We note here two stars whose results do not conform with rest of the sample.  When plotted in our \textit{preliminary} results, specifically Figures~\ref{fig:sodium}(right panel),~\ref{fig:alpha},~\ref{fig:br},~\ref{fig:spec} \&~\ref{fig:velocity}: \object{BD$+37\ 03856$} has an anomalous  spectral energy distribution (SED) for a \be star. Each of the other stars in our sample has a SED of the  the form $J\!>\!H\!>\!K$ or $K\!>\!H\!>J$, where $JHK$ are fluxes,  however the SED of this star is such that $J\!>\!H\!<\!K$.  This star also exhibited the most extreme point on our plots having the most negative $E(H-K)_{cs}$,  we therefore remove the point from our plots but for completeness list the  object in Table~\ref{tab:mean}. A possible explanation for this SED is thermal emission from dust, although 
we note that the object is not in the IRAS point source catalogue.

\object{BD$+57\ 00681$} also exhibits a large, negative $E(H-K)_{cs}$.  The random error on this object is small at $<1\%$ and it lies a long way from our calculated fit.  We find no reason however to remove the point from the data set.

\subsection{Br$\gamma$, Br11, Br18 and H$\alpha$}

\begin{figure*}
\vspace*{6.5cm}
\includegraphics{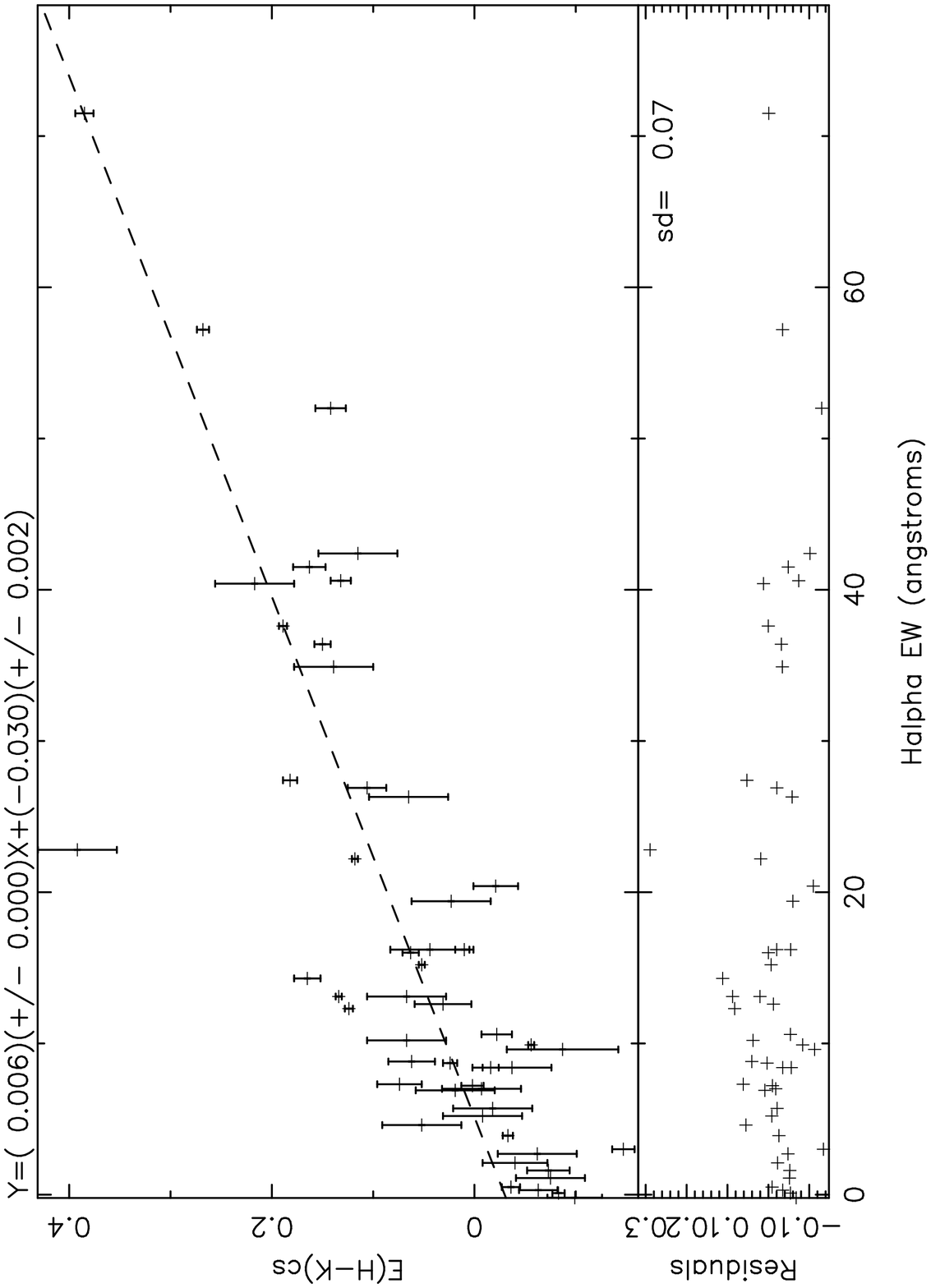}
\includegraphics{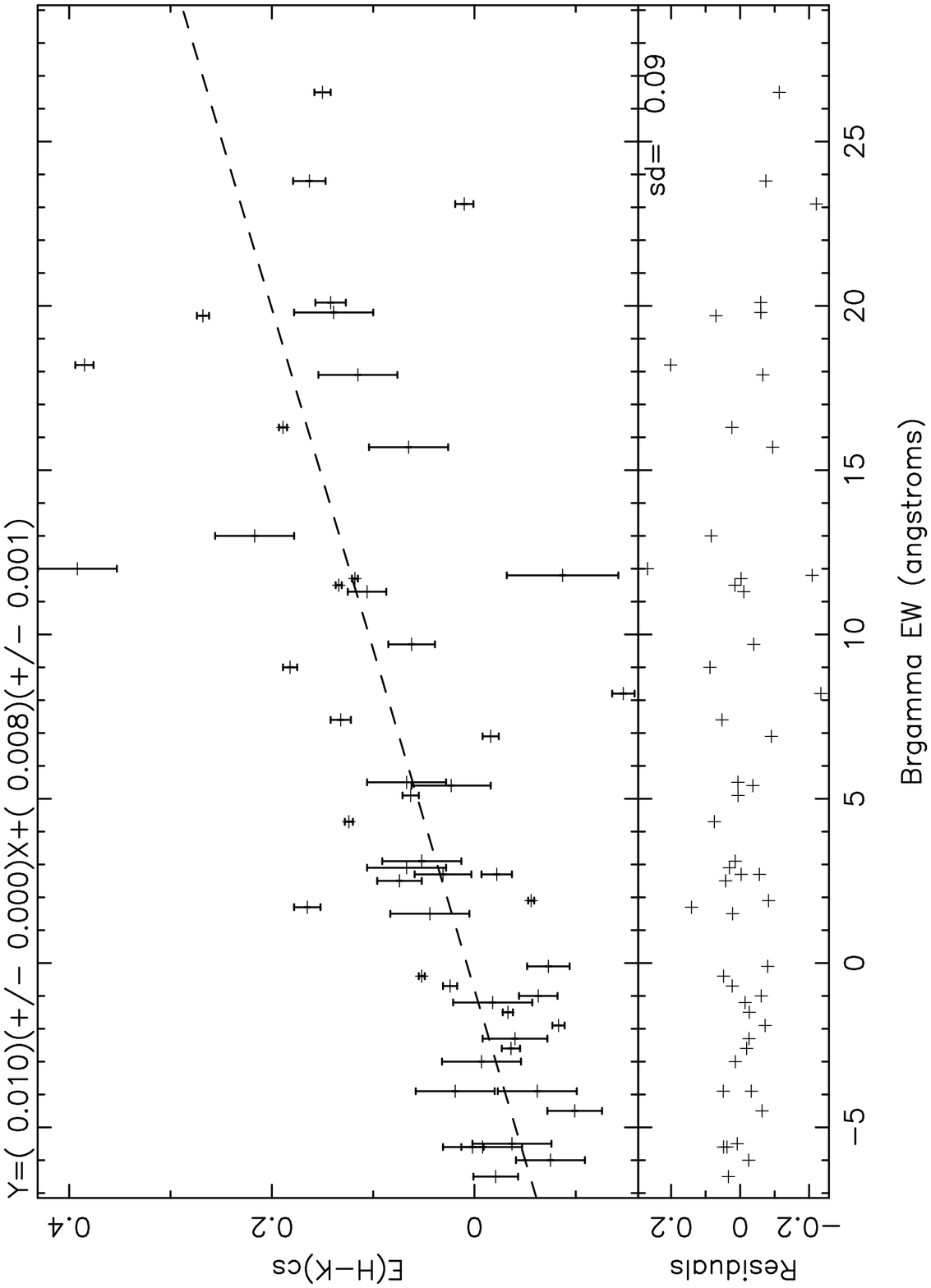}
 \caption{Circumstellar excess versus  $H\alpha$ EW/\AA (left panel) and  Br$\gamma$ EW/\AA (right panel).  The fitted lines  are least squares fits weighted to the ordinate axis errors}
\label{fig:alpha}
\end{figure*}

\begin{figure*}
\vspace*{6.5cm}
\includegraphics{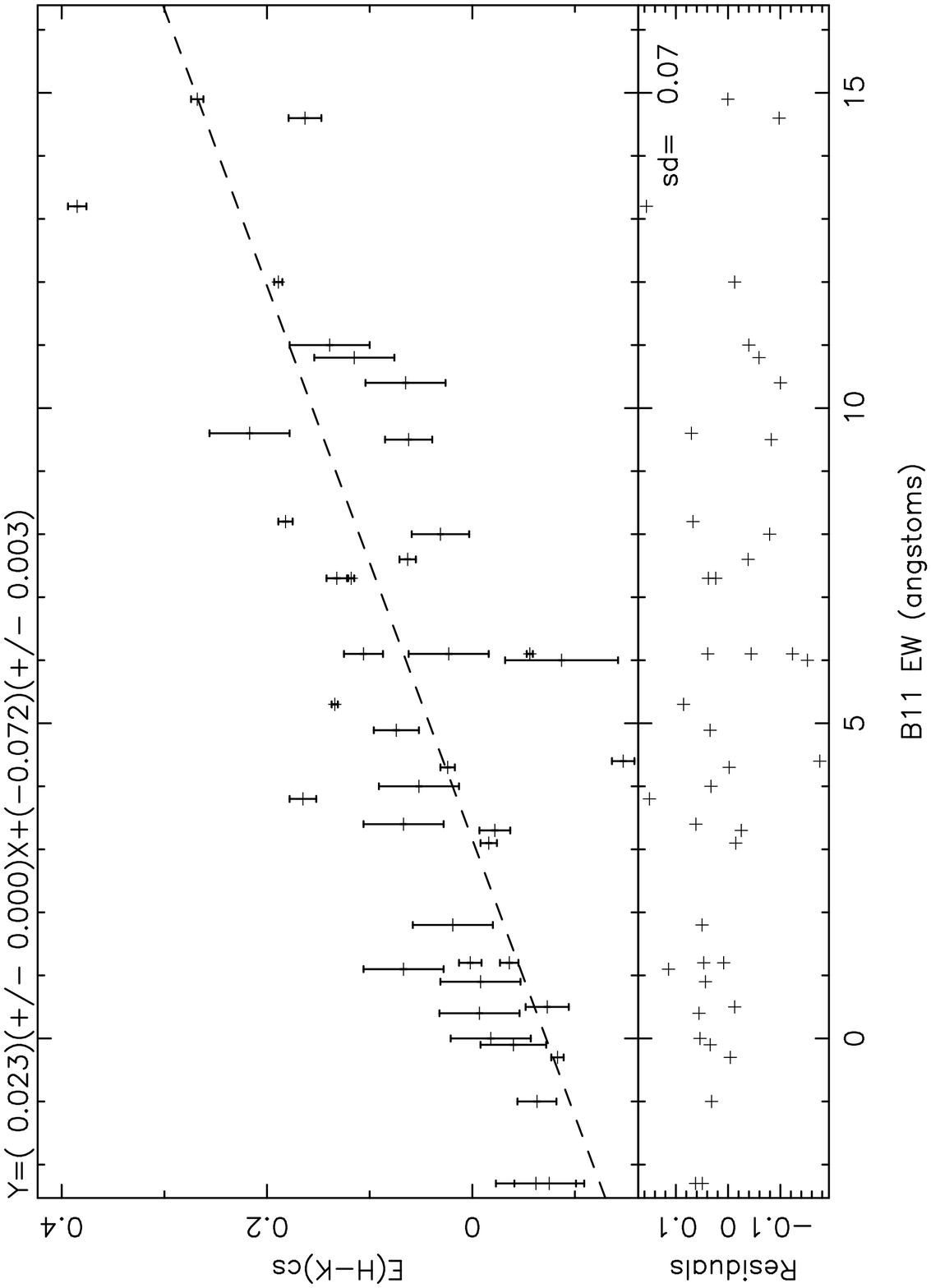}
\includegraphics{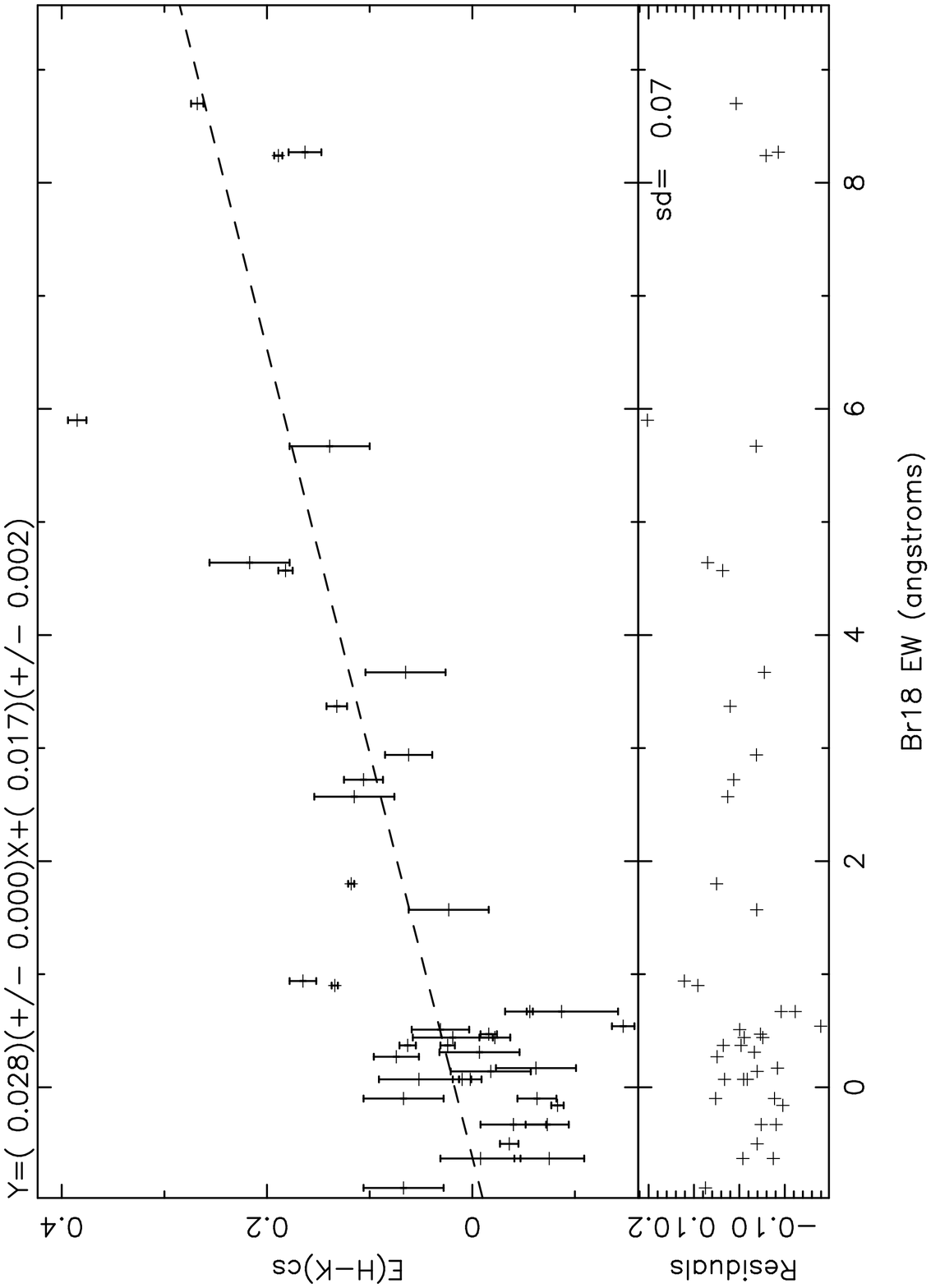}
\caption{Circumstellar excess versus Br11 EW/\AA (left panel)  and  Br18 EW/\AA (right panel). The fitted lines  are least squares fits weighted to the ordinate axis errors}
\label{fig:br}
\end{figure*}
Our Br$\gamma$ EW come from \cite{p2} and have an error of $\sim$10\%.
  Br18 EW and Br11 EW are extracted from \cite{p3} and also have errors of $\sim$10\%. 
 H$\alpha$ data come from \cite{p5} and again have an error of $\sim$10\%.

We plot the $E(H-K)_{cs}$ against Br$\gamma$, Br18, Br11 and H$\alpha$
in Figures~\ref{fig:alpha} and~\ref{fig:br}.  There is an obvious
correlation in each of the plots, re-enforced by the $>4.5\sigma$
confidence levels  produced  by the Spearman tests.    We fit lines of
least squares, weighted to the ordinate axis errors, to the data in
order to ascertain any linear correlation.

 It is worthy of note that van Kerkwijk \etal (\cite{k95}) present similar results to
ours   regarding the relationship  between line   equivalent width and
continuum excess, although they present H$\alpha$ versus J-L excess emission. 
In that study, as with our results there is a strong (apparently linear)
correlation between the lines and continuum excess. This correlation is
not surprising, as the hydrogen lines and the near-IR excess continuum are
typically formed in the same regions of the disc.  van Kerkwijk \etal (\cite{k95}) also
show the line-excess continuum correlations for two popular models of the
disc -- Waters' disc model (\cite{w86}), and the Poeckert \& Marlborough model (PM) (\cite{pm78}) --
and find that neither can replicate the results particularly well. The PM
model produces too little line emission for a given continuum excess, and
the disc model produces too much line emission, unless a large density
gradient is used (a radial density power law with an index larger than 3.5
seems to be necessary which appears to be inconsistent with the results
obtained from IRAS data). Whilst there is a large scatter
in the data, we present the linear best-fit relationship from our
results which any new model of Be star discs should attempt to reproduce.



\subsection{Helium I 2.058$\mu$m}
\begin{figure}
\vspace*{6.5cm}
\includegraphics{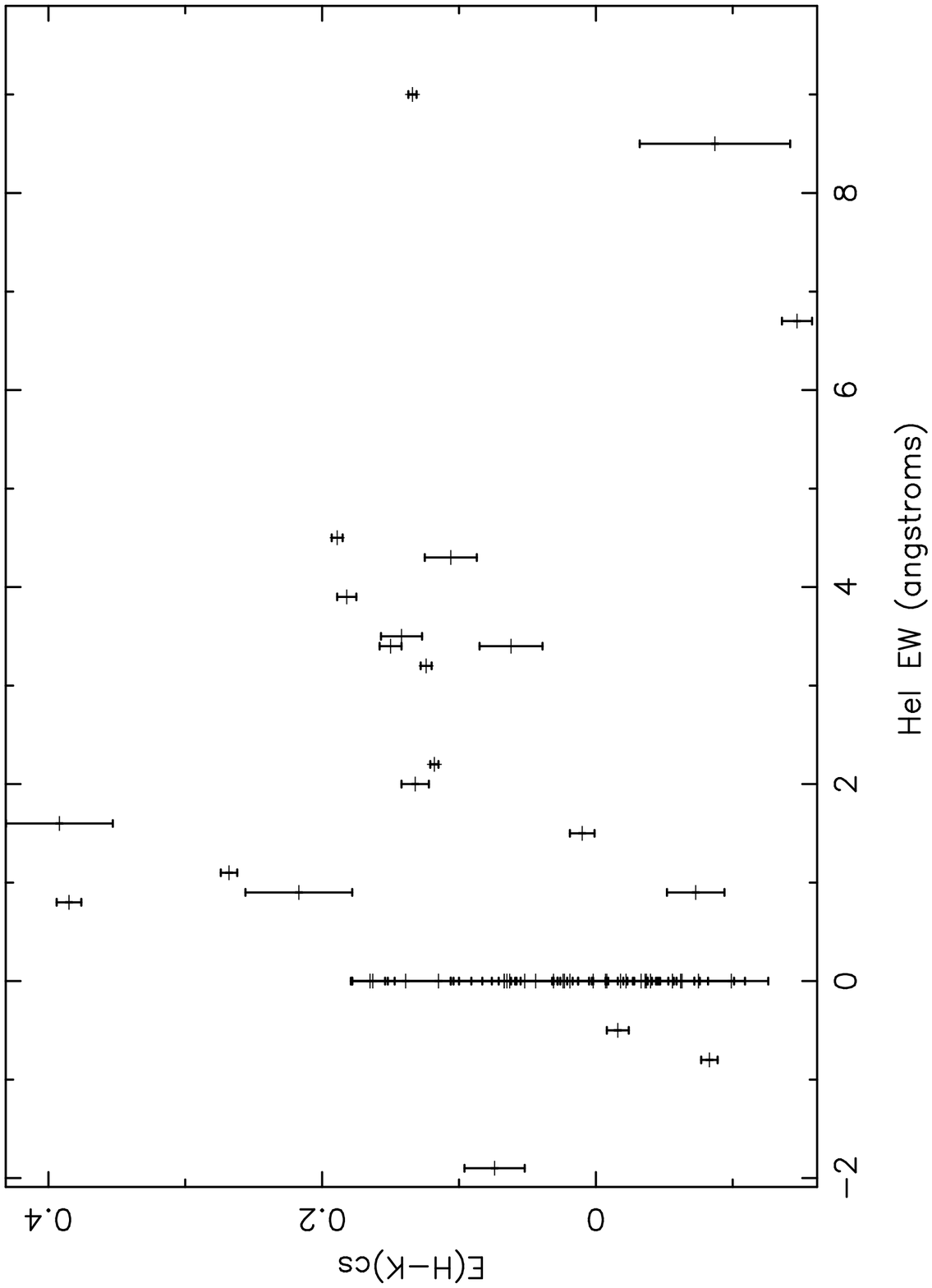}
\caption{Circumstellar excess versus HeI 2.058$\mu$m EW/\AA}
\label{fig:he}
\end{figure}

The HeI 2.058$\mu$m emission is confined to the early stars of the sample, being seen in 19 of the 34 stars with spectral types determined in \cite{p1} to be  earlier than B2.5. In Figure~\ref{fig:he} we plot $E(J-H)_{cs}$ versus HeI 2.058$\mu$m.  We note that data in Hanson \etal (\cite{h96}) shows the absorption lines for the HeI 2.0581$\mu$m line to be negligible and so no correction has been made.  Figure~\ref{fig:he} has $r=0.38$ and is therefore correlated at $>4\sigma$ confidence level, although no linear correlation seems to exist.  This is likely due to the fact that the HeI 2.058$\mu$m line is extremely sensitive to changes in the UV continuum and optical depth (\cite{p2}).  

\subsection{Spectral Type}
\begin{figure}[p]
\vspace*{6.5cm}
\includegraphics{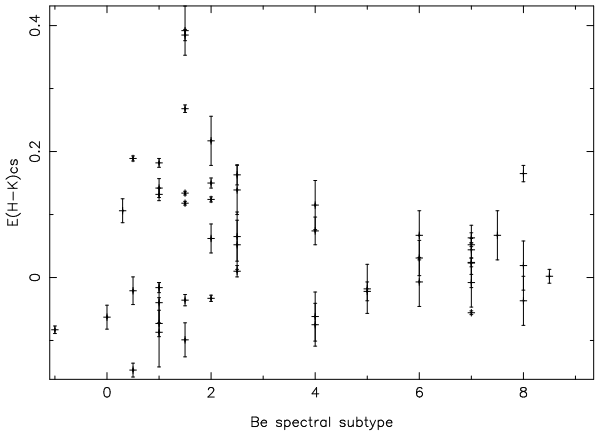}
\caption{Circumstellar excess versus spectral subtype}
\label{fig:spec}
\end{figure}

Figure \ref{fig:spec} shows a plot of circumstellar excess against spectral type, we note that there appears to be no linear correlation and that $r=4\times10^{-3}$ which gives a confidence level of  $>2\sigma$.  However, the overall shape of the distribution is similar to that seen in~\cite{p2} and~\cite{p3} for the strength of the Balmer series lines, with a broader range of excess around B1--B2. We note also that there is no correlation between luminosity class and circumstellar excess. 

\subsection{$v\sin(i)$  \& $\omega \sin(i)$}
\begin{figure*}
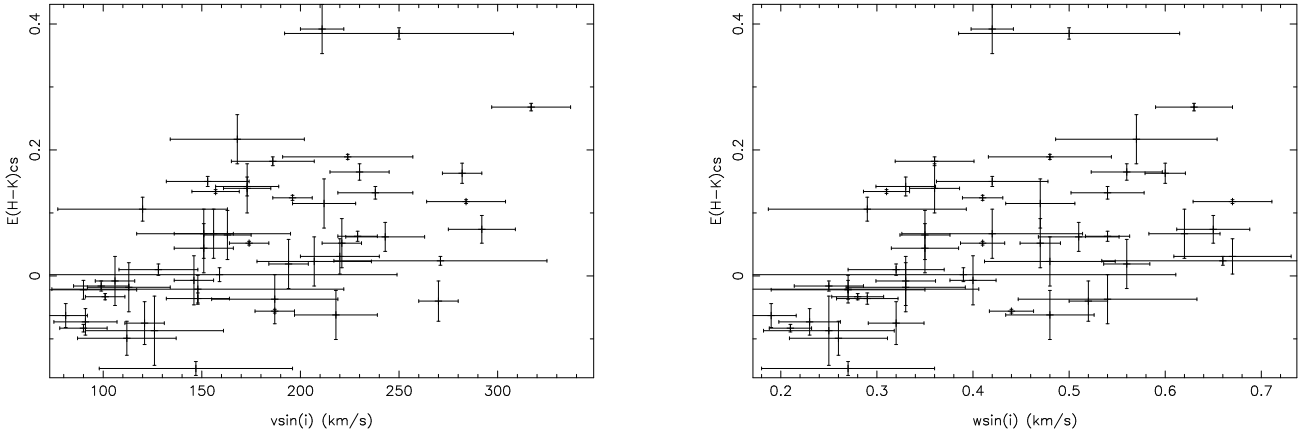

\vspace*{6.5cm}
\includegraphics{h2419f9a.ps}
\includegraphics{h2419f9b.ps}
\caption{Circumstellar excess versus $v\sin(i)$ (left panel) and $\omega\sin(i)$ (right panel). }
\label{fig:velocity}
\end{figure*}

Stellar rotation has been fundamentally linked with the generation of the \be phenomenon (\textit{e.g.} \cite{1988PASP..100..770S}, \cite{s82}). If rotation is the sole cause of the phenomenon then we would expect to see a strong correlation between rotational velocity and circumstellar excess. We therefore also plot $E(H-K)_{cs}$ versus $v\sin(i)$: see Figure~\ref{fig:velocity}, where velocity data are extracted from \cite{p1} and \cite{p2}.
Dougherty \etal (\cite{d94}), \cite{w86b} and Gehrz \etal (\cite{g74}) all find a similar result, that there is no correlation between $v\sin(i)$ and  colour excess.  However from Spearman rank tests we are able say that $v\sin(i)$ and $E(H-K)_{cs}$ (see Figure \ref{fig:velocity}, left panel) are related at a  $>4.5\sigma$ confidence level. In an attempt to remove spectral type dependence we also plot $\omega\sin(i)$ versus $E(H-K)_{cs}$, see Figure~\ref{fig:velocity}, right panel,  where $\omega\sin(i) = v\sin(i)/v_{crit}$ with $v_{crit}$ taken from \cite{p2}, (see Porter \cite{p96} for a discussion of the merits of using  $\omega \sin(i)$ compared to $v\sin(i)$).  While this plot exhibits a smaller scatter than our $v\sin(i)$ plot it is still correlated at confidence level of $>4.5\sigma$.  

\begin{table*}
\begin{tabular}{lcccccc}
\multicolumn{1}{c}{Plot}&
\multicolumn{1}{c}{$r$}&
\multicolumn{1}{c}{Sig.\ level}&
\multicolumn{1}{c}{Conf. level}&
\multicolumn{1}{c}{Stan. dev.}&
\multicolumn{1}{c}{Gradient}&
\multicolumn{1}{c}{Intercept}\\ 
\hline 
\\
$E(H-K)_{cs}$ vs $H\alpha$ 		&0.83    &  $0.005$ &$>4.5\sigma$  &$\pm$0.07 & 0.006$\pm$0.000 &-0.030$\pm$0.002\\
$E(H-K)_{cs}$ vs Br11                   & 0.75   &  $0.005$ &$>4.5\sigma$  &$\pm$0.07 & 0.023$\pm$0.000 &-0.072$\pm$0.003\\
$E(B-V)_{{(H-K)}_{is}}$ vs $E(B-V)_{is+cs}$ &0.74&  $0.005$ &$>4.5\sigma$  &$\pm$0.25 & 0.814$\pm$0.007 &0.168$\pm$0.003\\ 
$E(H-K)_{cs}$ vs Br18                   & 0.70   &  $0.005$ &$>4.5\sigma$  &$\pm$0.07 & 0.028$\pm$0.000 &0.017$\pm$0.002\\
$E(H-K)_{cs}$ vs Br$\gamma$             & 0.69   &  $0.005$ &$>4.5\sigma$  &$\pm$0.09 & 0.010$\pm$0.000 &0.008$\pm$0.001\\
$E(H-K)_{cs}$ vs $v\sin (i)$            & 0.58   &  $0.005$ &$>4.5\sigma$  & -   & -             & -\\
$E(H-K)_{cs}$ vs $\omega\sin (i)$       & 0.54   &  $0.005$ &$>4.5\sigma$  & -   & -             & -\\
$E(B-V)_{is+cs}$ vs  sodium EW          & 0.56   &  $0.005$ &$>4.5\sigma$  &$\pm$0.24 & 0.39 & 0.13 \\
$E(H-K)_{is}$ vs sodium EW              & 0.47   &  $0.005$ &$>4.5\sigma$  &$\pm$0.06 &0.065$\pm$0.005 &0.049$\pm$0.003\\
$E(B-V)_{{(H-K)}_{is}}$ vs sodium EW 	& 0.45	 &  $0.005$ &$>4.5\sigma$  &$\pm$0.31 & 0.37 & 0.15 \\
$E(H-K)_{cs}$ vs HeI EW                 & 0.38   &  $0.37$  &$>4\sigma$  & -  & -         & -\\
$E(H-K)_{cs}$ vs spectral type  &$4\times10^{-3}$&  $ -   $  &$>2\sigma$  & -  & -  & -\\
$E(H-K)_{cs}$ vs sodium EW              & -0.08  &  $ -   $  &$<1\sigma$  & -  & -         & -\\
\hline
\end{tabular}
\caption{\label{tab:spear} We present our data table in order of descending Spearman rank coefficient, $r$ (Col. 2), Col 3, sig. level, are the one-tailed significance levels of the Spearman rank Correlation and are extracted  from \cite{1996QJRAS..37..519W}. Col 4 are the one-tailed confidence levels of our correlations and are extracted from  \cite{1979QJRAS..20..138W} and Col. 5 is the standard deviation about the fitted least squares fits line in the ordinate direction.  In Col. 6 we present the gradient of those fits and Col. 7 is the intercept.}
\end{table*}

\section{Conclusions}
\label{sec:conc}
We have presented 
IR ($JHK$) photometry of the ``representative sample'' of Be stars defined by ~\cite{p1}. We have derived a new technique for separating the effects of (i) emission of the circumstellar material, and (ii) interstellar reddening.  The technique involves combining photometry of 3 IR filters, a general  interstellar extinction law and an assumption (that we verify) that the colours of the circumstellar disc of a \be star can be related in a similar fashion. 

By correlating our results with an independent measure of interstellar reddening (the equivalent width of the Na I 5890\AA \ line), we are confident that our method is valid.  Using this technique we find that the disc emission makes a maximum contribution to the optical $(B-V)$ colour of a few tenths of a magnitude. 

We find significant correlations between the infrared excess emission from the disc and emission from a range of lines (H$\alpha$, Br$\gamma$, Br11, and Br18). Linear fits to these correlations have been derived. There is also a significant correlation between the $v\sin(i)$ (and also  $\omega\sin(i)$) of the Be star and the infrared disc emission.  These data present strong constraints for models of the structure of \be stars. Any model that purports to explain their disc structure must be able to reproduce the correlations presented in Figures~\ref{fig:alpha}--\ref{fig:velocity}.  In a future study we will contrast these observations with predictions from current models.

\begin{acknowledgements}
The TCS is operated on the island of Tenerife by the Instituto de Astrofisica de Canarias. Thanks to Chris Collins for useful discussions relating to the statistics of this paper. LH greatfully acknowledges PPARC funding. 
\end{acknowledgements}


\begin{thebibliography}{}
\bibitem[1998]{a98} 
Alonso A., Arribas S.  Martinez-Roger, C. 1998, A\&AS 131, 209 
\bibitem[1993]{bc93}
Bjorkman J. E. \& Cassinelli J. P. 1993, ApJ 409, 429
\bibitem[1984]{c84} 
Cramer N. 1984, A\&A 132, 283 
\bibitem[Paper II]{p2}
Clark J.S., Steele I.A., 2000, A\&AS 141, 65 -  Paper II
\bibitem[1988]{d88} 
Dachs J., Kiehling R. Engels D. 1988, A\&A 194, 167 
\bibitem[1986]{d86} 
Dachs J., Hanuschik R., Kaiser D.\ Rohe D.\ 1986, A\&A 159, 276 
\bibitem[1994]{d94}
Dougherty S.M., Waters L.B.F.M., Burki G., Cote J., Cramer N., van Kerkwijk M.H., Taylor A.R., 1994, A\&A 290, 609
\bibitem[Dougherty \& Taylor 1992]{1992Natur.359..808D} Dougherty S.\ 
M.\  Taylor A.\ R.\ 1992, Nature 359, 808 
\bibitem[1982]{e82} 
Elias J.\ H., Frogel J.\ A., Matthews K.\ Neugebauer G.\ 1982, AJ 87, 1029 
\bibitem[1974]{g74}
Gehrz R.D., Hackwell J.A., Jones T.W., 1974, ApJ 191, 675
\bibitem[Hirata 1995]{1995PASJ...47..195H} 
Hirata R.\ 1995, PASJ 47, 195 
\bibitem[1996]{h96}
Hanson M.M., Conti P.S., Rieke, M.J., 1996, ApJs 107, 281
\bibitem[1998]{h98}
Hanson M.M., Rieke G.H.,  Luhman K.L., 1998, AJ 116, 1915
\bibitem[Hanuschik \etal 1993]{1993A&A...274..356H} 
Hanuschik R.\ W., Dachs J., Baudzus M.\ Thimm G.\ 1993, A\&A 274, 
356
\bibitem[2000]{hr2000}
Hummel W.  Vrancken M., 2000, A\&A 359, 1075 
\bibitem[1995]{k95} 
van Kerkwijk M.\ H., Waters L.B.F.M.,  Marlborough J.\ M.\ 
1995, A\&A 300, 259 
\bibitem[1989]{kast89}
Kastner J.H.  Mazzali, 1989, A\&A 210, 295 
\bibitem[1983]{k83}
Koornneef J., 1983, A\&AS 128, 84
\bibitem[Lamers \& Pauldrach 1991]{1991A&A...244L...5L} 
Lamers H.J.G.,  Pauldrach A.W.A., 1991, A\&A 244, L5 
\bibitem[1991]{1991MNRAS.250..432L} 
Lee U., Osaki Y.,  Saio, H., 1991, MNRAS 250, 432 
\bibitem[1992]{leg92} 
Leggett S. K. 1992, ApJ 82, 351

\bibitem[Okazaki 1997]{1997A&A...318..548O} 
Okazaki A.T., 1997, A\&A  318, 548

\bibitem[Osaki 1986]{1986PASP...98...30O} Osaki Y., 1986, PASP 98, 30 
\bibitem[Owocki \etal, 1994]{1994ApJ...424..887O} 
Owocki S.P., Cranmer S.R., Blondin J.M., 1994, APJ 424, 887 
\bibitem[1978]{pm78}
Poeckert R., Marlborough J.M., 1978 ApJ 220, 940
\bibitem[1996]{p96}
Porter J.M., 1996, MNRAS 280, L31
\bibitem[1999]{1999A&A...348..512P} 
Porter J.M., 1999, A\&A 348, 512 
\bibitem[1992]{spearman}
Press W.H., Teukolsky S.A., Vetterling W.T., Flannery B.P., 1992, Numerical Recipes in C, The Art of Scientific Computing, 2nd edn.\ Cambridge Univ.\ Press.
\bibitem[Quirrenbach \etal 1994]{1994A&A...283L..13Q} Quirrenbach A., 
Buscher D.F., Mozurkewich D., Hummel C.A.,  Armstrong J.T., 
1994, A\&A 283, L13 
\bibitem[1985]{rb85}
Rieke G.H., Lebofsky M.J., 1985, ApJ 288, 618
\bibitem[1982]{s82}
Slettebak A., 1982, ApJS 50, 55
\bibitem[Slettebak 1988]{1988PASP..100..770S} 
Slettebak, A., 1988, PASP 100, 770 
\bibitem[Paper I]{p1}
Steele I.A., Negueruela I., Clark J.S., 1999, A\&AS 137, 147 - Paper I
\bibitem[Paper III]{p3}
Steele I.A., Clark J.S., 2000, A\&AS in press - Paper III
\bibitem[Paper V]{p5}
Steele I.A., Negueruela I., A\&AS in prep., - Paper V
\bibitem[Waters (1986b)]{w86b} 
Waters L.B.F.M., 1986b, A\&A 159, L1 
\bibitem[1986a]{w86}
Waters L.B.F.M., 1986a, A\&A 162, 121
\bibitem[Wall (1979)]{1979QJRAS..20..138W} 
Wall J.V., 1979, QJRAS 20, 138 
\bibitem[Wall (1996)]{1996QJRAS..37..519W} 
Wall J.V., 1996, QJRAS 37, 519 

\end{thebibliography}
\end{document}